# Magnetic anomalies in $Nd_6Co_{1.67}Si_3$: Surprising first order transitions in the low-temperature isothermal magnetization

Niharika Mohapatra, S. Narayana Jammalamadaka, Sitikantha D. Das and E.V. Sampathkumaran[*]

*Tata Institute of Fundamental Research, Homi Bhabha Road, Colaba, Mumbai – 400005, India*

Abstract

We present the results of magnetic measurements on $Nd_6Co_{1.67}Si_3$, a compound recently reported to crystallize in a hexagonal structure (space group, $P6_3/m$) and to undergo long range magnetic ordering below 84 K. The results reveal that the magnetism of this compound is quite complex with additional magnetic anomalies near 50 and 20 K. There are qualitative changes in the isothermal magnetization behavior with the variation of temperature. Notably, there is a field-induced spin reorientation as the temperature is lowered below 20 K. A finding we stress is that this transition is discontinuous for 1.8 K in the virgin-curve, but the first-order character appears *only after a field-cycling* for a narrow higher temperature range near 5 K. Thus, this compound serves as an example for the stabilization of first-order transition induced by magnetic-field cycling. The issues 'phase co-existence' and 'metastability' after a field-cycling at low temperatures in this compound are also addressed.

## I. INTRODUCTION

The compounds exhibiting first-order magnetic transitions induced by magnetic-field have been of great current interest. In particular, the rare-earth intermetallics with this behavior have been of great value from the angle of magnetic phase coexistence phenomenon. In this respect, $CeFe_2$-derived alloys [1], $Gd_5Si_4$ [2], and $Nd_7Rh_3$ [3], attracted some interest. In general, such first-order transitions may be broadened due to static effects like crystallographic disorder. In this article, we present the results of our magnetic investigations on the compound, $Nd_6Co_{1.67}Si_3$, recently reported to order magnetically below 84 K [4, 5]. The results, apart from establishing the existence of additional magnetic anomalies near 50 and 20 K, bring out a feature presumably unique in the field of first-order magnetic transitions. We find that there are spin re-orientation transitions induced by magnetic field (H) appearing gradually with a decrease of temperature across 20 K. Such transitions appear to be first-order in the virgin curve for 1.8 K. For a marginal increase of temperature, say to 5 K, this transition in the virgin curve is broadened presumably due to disorder. The unique observation is that these continuous transitions could be made first-order after a field-cycling.

## II. CRSYSTALLOGRAPHIC DETAILS

The family of $R_6Co_{1.67}Si_3$ (R= Rare-earths) has been reported to form only very recently [4, 5] with a hexagonal structure derived from $Ce_6Ni_2Si_3$ (space group: $P6_3/m$) [6,7]. The structure contains chains of trigonal prisms of Nd filled by Co and each face of this trigonal prism is shared by chains of Si-centered Nd-trigonal prisms along the basal plane, as shown in figure 1a. These four chains thus run along $c$-axis. Columns of Nd octahedra sharing the faces among themselves along $c$-axis as shown in figure 1b bridge these triangular columns. In these octahedra, a part of Co occupies a position which is slightly displaced from the center, and this site is randomly occupied due to a strain resulting from small $c$ parameter. This accounts for Co stoichiometry less than 2. The shared faces of the octahedra are also preferred by Co due to this reason. Needless to state that, from the above crystallographic description (see also figure 5 of Ref. 10), there are two sites for R, 3 sites for Co and one site for Si.

## III. EXPERIMENTAL DETAILS

The polycrystalline specimen of the compound under investigation has been prepared by arc-melting stoichiometric amounts of high purity (>99.9%) constituent elements together in an atmosphere of argon. The weight loss after several melting was less than 0.5%. The ingot thus obtained was annealed at 1073 K for 1 month in an evacuated sealed quartz tube. The sample was characterized by x-ray diffraction (Cu Kα) and the pattern reveals the formation of the compound, though a few weak extra lines were observed due to a secondary phase (a few percent) as known for this family in the past [8]. The lattice constants (±0.004 Å) ($a$= 11.96 Å and $c$= 4.25 Å) are found to be in good agreement with those reported in the literature [5]. The temperature (T) dependent (1.8 - 300 K) dc magnetization (M) measurements were performed with a commercial vibrating sample magnetometer (VSM) (Oxford Instruments) in the presence of 100 Oe, 500 Oe and 5 kOe for the zero-field-cooled (ZFC) and field-cooled (FC) conditions of the specimen in the form of an ingot; in addition, isothermal M data presented here for several temperatures were measured below 120 kOe while sweeping the magnetic field with a rate of 4 kOe/min (unless otherwise stated). A commercial SQUID magnetometer (Quantum Design) was employed to perform ac susceptibility ($\chi_{ac}$) measurements (1.8 – 120 K) with an ac field of 1 Oe at different frequencies (1.3, 13, 133, and 1333 Hz); in addition certain magnetic relaxation and isothermal M studies were also carried out with the same magnetometer. Heat-capacity (C) measurements were performed below 140 K in zero-field and in 40 kOe employing a commercial physical property measurements system (Quantum Design).

## IV. RESULTS AND DISCUSSION
### A. Temperature dependence of dc magnetization

We show in figure 2 the T-dependence of dc M measured in the presence of various fields. Curie-Weiss behavior of dc $\chi$ is observed above 90 K with a value of the effective moment (3.60 $\mu_B$/Nd) close to that expected for trivalent Nd ions, thereby implying that Co does not carry any magnetic moment in the paramagnetic state. The paramagnetic Curie temperature obtained from the linear region turns out to be about 80 K, the positive sign of which indicates dominance of ferromagnetic coupling. These results are in good agreement with those reported in the literature [4 and 5]. In the data taken with low fields (say, 100 Oe), there is a sudden increase of M near 84 K due to the onset of long-range magnetic order. The transition width is large in the data taken with 5 kOe. Below 84 K, there are qualitative differences among the shapes of the curves obtained with different strengths of the applied field and the cooling condition of the specimen, in broad agreement with Refs. 4 and 5. For instance, ZFC and FC curves bifurcate (due to domain wall effects) and the temperature at which this irreversibility sets in depends upon the field, for instance, near 84 K for H= 100 Oe, 75 K for H= 500 Oe, and 20 K for H= 5 kOe. It appears that there is another magnetic anomaly near 50 K, which undergoes profound changes with the application of magnetic fields. We will offer further evidence for the same from ac $\chi$ data later in this article. This anomaly possibly manifests in the form of a peak in ZFC-$\chi$(T) curve and a knee in FC-$\chi$ curve in the range 42-45 K, say, for H= 500 Oe. Similar features were reported in Ref. 4, but at a somewhat lower temperature (38 K). Thus, the overall magnetism appears to be complex. It is not clear whether this complexity arises due to two different types of Nd ions and/or due to induced-magnetism on Co influencing spin orientations.

### B. Isothermal magnetization

We now focus on the M(H) behavior (see figure 3), taken at close intervals of temperature. Since there is an irreversibility in the ZFC-FC M(T) curves, there is a need to warm up the sample to the paramagnetic state and to cool in zero-field to the desired temperature before collecting data at any temperature. We have performed three sets of independent measurements for this virgin state for each temperature with VSM: (i) M behavior with the variation of H from 0 to 60 to 0 kOe (see figure 3a and 3b for the data below 35 kOe), though for a few temperatures, the measurements were extended to 120 kOe; (ii) hysteresis-loop behavior for a variation of H from 0 to 30 to -30 to 30 to 0 kOe (figure 4); (iii) hysteresis-loop behavior in the range ± 2 kOe (figures 3c and 3d). The density of data points is so large that the plots appear as continuous lines in these figures. It is found that the M is hysteretic at all temperatures in the magnetically ordered state. As shown in figure 4 for 60, 20, 10 and 5 K, the size of the hysteresis-loop, if measured over an extended field range, increases gradually with decreasing temperature down to 5 K.

A closer look at the M(H) behavior is quite revealing. The main point of emphasis is that the M(H) behavior appears to sensitively depend on the temperature as well as on the field-cycling as follows:

- For 80K>T>20K, there is a sharp increase of M below about 2 kOe following which there is a retarded and gradual variation of M with H (see figure 3b), without saturation even at fields as high as 120 kOe (not shown in the figure as there is no other feature).

The saturation moment obtained by linear extrapolation of the high-field data (even at 1.8 K, 2.3 $\mu_B$/Nd) is far below that expected for fully-degenerate trivalent Nd ions, which could imply the existence of an antiferromagnetic component, perhaps including canting as well as crystal-field effects. For 20 K, additionally, there is a weak step followed by a dramatic rise in the range of 2 - 5 kOe, superimposed over this initial rise below 2 kOe, as though there is a spin-reorientation (see figure 3a). This step near 2 kOe persists down to 1.8 K, but gets gradually more prominent with a decrease in temperature below 20 K (see figure 3a) with the spin-reorientation feature shifting to a higher field. For instance, this spin-reorientation manifests as a broad transition around 22 kOe at 5 K. Thus, around 20 K, the magnetic behavior undergoes a peculiar change.

- The spin-reorientation transition appears as a first-order transition (in the virgin-curve itself) around 26 kOe for T= 3 K and, interestingly, at a lower field (near 20 kOe) for 1.8 K.
- This spin re-orientation transition does not appear when the field is reversed towards zero. This means that these transitions are irreversible.
- The most intriguing finding is that, at 5 K, if M is measured while increasing the field after reversing the current direction to the magnet (that is, in the hysteresis loop, path-3 in figure 4a), instead of a broad transition around -20 kOe similar to the one in the virgin curve, there is a sharp transition near -22 kOe; for a further increase of H to 30 kOe (from -30 kOe along path-4), *the broad transition appearing in the virgin curve is replaced by a sharp transition* near 19 kOe. At this juncture, it is worth mentioning that, in $Nd_7Rh_3$, there is a field-induced first-order magnetic transition in the virgin curve below 10 K, which is broadened after a field-cycling [3]. The observation in the present compound is interestingly opposite to this behavior. Thus, the present compound serves as an example for a situation in which field-cycling stabilizes a first-order transition, which is somehow broadened (presumably by disorder) in the virgin curve. It is also interesting to note that *a significant part* of the virgin curve is outside of the subsequent M(H) hysteresis envelope for T≤5K, whereas in $Nd_7Rh_3$, entire virgin curve lies outside envelope curve at low temperatures. As the temperature is raised, say to 10 K, the spin-reorientation transition remains as a broad feature; however, after the field-cycling, it occurs at a marginally higher field (near 15 kOe), while compared to that in the virgin curve (near 10 kOe, see figure 4a); the virgin curve lies well-inside the envelope curve in contrast to the behavior at lower temperatures..

We now turn to low-field M(H) data shown in figures 3c and 3d for some representative temperatures for ZFC condition of the specimen to infer the nature of magnetic structure with varying temperature. We restrict these studies to ± 2 kOe to avoid possible interference from spin-reorientation effects discussed above and to explore whether the high-field hysteresis-loop below 20 K shown in figure 4 is superimposed over another low-field loop. It is clear that M(H) is hysteretic at all temperatures. In these loops, the virgin curve lies inside the envelope curve in contrast to that observed in the hysteresis loops extended to high-fields (figure 4) even at 1.8 K, which suggests that the opposite behavior observed in the latter case is the result of spin-flip transitions. The observed hysteresis in figures 3c and 3d reveal the presence of a ferromagnetic component down to 1.8K. Needless to state that there is an antiferromagnetic component as well, as one observes spin-orientation effects at higher fields. The presence of both ferro- and antiferromagnetic component prompts one to infer that the ZFC cooled state at low temperatures could be a spin-glass, a possibility that can not be excluded due to crystallographic disorder at a Co site discussed in Section II. However, ac $\chi$ data presented later in this article is against this possibility. These experimental facts lead us to believe at present that the zero-field state prior to any high-field cycling may in fact be described as a 'ferrimagnet' at all temperatures. It is not clear how ferrimagnetism arises and how these magnetic structures differ in different temperature ranges. To facilitate ferrimagnetism, it is possible that the magnetic moment on Nd at the two sites are not the same due to different crystal-field-split ground state arising from different crystallographic environment. Additionally, moment on Co can be induced by magnetically ordered Nd ions. We found an evidence for the induced moment (> 1 $\mu_B$ per Co ion) on Co on the basis of our (unpublished) results on analogous Gd compound. These moments can couple antiparallel to each other.

### C. Ac susceptibility

The results of $\chi_{ac}$ studies are shown in figure 5. We have obtained the data for three situations: (i) After ZFC to 1.8 K, the data were collected; (ii) After ZFC to 1.8 K, a magnetic field of 30 kOe was switched on and the data were collected; and (iii) ZFC to 1.8K, switch on a field of 30 kOe and the data were taken after switching off the field. In each case the curves were obtained while warming.

We note that, for (i), in addition to an intense peak at 84 K due to the onset of magnetic ordering, an additional sharp peak near 50 K is observed in both real and imaginary parts supporting the existence of another magnetic anomaly inferred from dc magnetization data (figure 2). A notable point is that there is no frequency dependence of the peak positions within 0.1K. This is taken as a key evidence against spin-glass freezing. An additional feature is that there is a gradual decrease below 30 K, as evident in the real part of $\chi_{ac}$ in figure 5, indicative of another magnetic anomaly in the vicinity of this temperature. In the same temperature range, the features in the dc M data (Section IV-B) revealed a magnetic anomaly. Considering that the feature in $\chi_{ac}$ is very broad around 30 K, the anomalies near 20-30 K may not be strictly called 'a magnetic transition'. Possibly, there is a continuous spin-orientation with decreasing temperature. Further microscopic studies are required to understand this interesting anomaly.

With respect to the data measured for the situation (ii), the features were completely washed out and this appears to be a characteristic of high-field 'ferromagnetic state'. For the measurement conditions as specified in (iii), the virgin state (zero-field) features appear at the same temperatures, however with a modification of relative intensities of the peaks. A notable observation is that the signal is extremely weak (only about 2% of the intensity of the virgin specimen) as shown in figure 5 as though the specimen in fact has only partially attained the magnetism corresponding to the virgin state. A combined look at the features for situations (ii) and (iii) leads us to the inference that there is a 'remanence' of the ferromagnetic state after reducing the field to zero. In what follows, we will offer evidence for a small decay of this ferromagnetic phase and therefore the question rises whether this 'mixed-state' can be described as a 'spin-glass' phase. The fact that there is no frequency dependence of the peak-positions, even after field-cycling implies that this is not the case. Thus, the mixed-phase is different from spin-glasses as argued earlier by us [3] and the term 'magnetic glass' has been used to describe such a phase [1].

### D. Time dependent isothermal magnetization

In order to address the issue of 'metastability', we have performed magnetic relaxation studies employing the SQUID magnetometer in the presence of fields at 1.8 and 7 K at different points in the M(H) curve in figure 6 along the paths 1 and 2, e.g., 0, 1, 10, 18, and 30 kOe. For H= 0, 1, 10 and 18 kOe, the data were collected in the reverse cycle as well. We ensured that, before attaining requisite field and temperature, the specimen was warmed up to 120 K (to paramagnetic state) and cooled in zero-field every time. We found that the magnetization does not decay with time in all these cases, except that, *in the reverse cycle, for H= 0 and 1 kOe*, we find a distinct evidence for a weak decay of M by about 1% (or less) in about an hour. The magnitude of M varies logarithmically with time (after a slower decay for initial few minutes) as shown in figure 6. Such a relaxation can not be interpreted in terms of spin-glass freezing, as ac $\chi$ data discussed above rules out this possibility. We therefore interpret that, in the reverse cycle, for H>>1 kOe, the high-field ferromagnetic phase is stable and, and for H= 0 and 1 kOe, high-field ferromagnetic and virgin-state-like magnetic phases coexist (following relaxation and kinetic arrest [1]).

### E. Heat capacity

The C(T) behavior of this compound in zero field and in 40 kOe are shown in figure 7. Though, in the ac $\chi$ data, there are additional peaks/drops near 50 and 20 K in support of more magnetic anomalies as the temperature is lowered, there is no evidence for any additional peak at corresponding temperatures in C(T) curve. This indicates that the entropy associated with these features must be small. We ignore an interpretation in terms of spin-glass freezing due to ac $\chi$ behavior described above. In order to address this aspect more carefully, we have also obtained C(T) curve in the presence of 40 kOe and derived isothermal entropy change ($\Delta S$) and adiabatic temperature change ($\Delta T$), since, qualitatively speaking, such magnetocaloric (MCE) properties have been known to yield information about magnetism [11,12]. For comparison, we have obtained $\Delta S$ from isothermal M behavior measured at close intervals of temperature (some of which are shown in figure 3b) in light of well-known Maxwell's relationship between $\Delta S$ and M. The results of $\Delta S$ thus obtained are shown in an inset in figure 7 for a change of the field from zero to a desired value. It is found that there is a good agreement between the $\Delta S$ behavior obtained from M and C data above 20 K. The values peak at the onset of magnetic ordering. The sign of $\Delta S$ is negative above 20 K, which is typical of systems with a dominating ferromagnetic component. It is evident that additional shoulders both in $\Delta S$ and $\Delta T$ around 50 K are present. Possibly, there is a subtle spin reorientation occurring at this temperature as suggested in Ref. 4. However, there is a sign reversal in the $\Delta S$ data obtained from M below 20 K with a significant magnitude, whereas the values obtained from C(T) are negligibly small. We attribute this discrepancy to profound sensitivity of magnetization to a small thermal/field cycling effects below this temperature, thereby vitiating a meaningful comparison with the $\Delta S$ derived from C(T) data at such low temperatures. In any case, this MCE discrepancy below 20 K endorses our earlier inference that this is a characteristic temperature at which there is a qualitative change in the magnetic behavior of this compound.

### V. Summary

We have reported the results of detailed dc and ac magnetization and heat capacity studies on the compound, $Nd_6Co_{1.67}Si_3$. Magnetic ordering sets in below 84 K in this compound. However, there are additional magnetic anomalies near 50 K and 20 K; though the entropy associated with these anomalies is small, the magnetocaloric properties render support to the existence of these features. Possibly temperature-induced spin-orientation effects result in these anomalies in a subtle way. Isothermal dc magnetization also undergoes complex changes with decreasing temperature. A notable finding in dc magnetization data is that there is a metamagnetic transition appearing at very low temperatures only, say, at 1.8 K, for the virgin state of the specimen. This transition is broadened in the virgin curve for a small increase of temperature, but appears after a field-cycling. In our opinion, this is an unusual observation in the field of metamagnetic transitions, as the behavior seems to imply the stabilization of the first-order transition by some unknown 'feed-back' mechanism induced by high-field cycling. In the literature, conventionally, crystallographic disorder has been known to smear out the first-order transitions, but such static effects alone can not explain the appearance of abrupt transitions after a field-cycling. In addition, there are features attributable to phase-coexistence and metastability after a field-cycling at very low temperatures. In short, this compound exhibits many magnetic anomalies. In view of this, there is an urgent need to perform exhaustive neutron diffraction studies as a function of temperature and magnetic field on this compound.

Finally, we would like to make some remarks on a few fascinating findings we have observed: (i) We repeated the studies on a few other specimens. The observations - in particular the absence of a jump in the virgin curve and its appearance after a field-cycling for T>1.8 K - made in this article are found to be reproduceable. However, the transition field can be slightly different for different specimens, and, in one specimen, another jump in M(H) curve for 1.8 K while increasing the field can be seen. We believe that this sample dependence may be intimately connected to the disorder at a Co site briefed in Section II. (ii) There is a negligible influence of rate of cooling (as measured with SQUID magnetometer for 10K/min and 3K/min) on the first-order transitions; but, there is a small dependence (<10 kOe) of the transition-field to the rate of change of field (as measured with VSM, 1 to 4 kOe/min) as in $Nd_7Rh_3$ [3]. We hope that these observations, apart from revealing complexities associated with the magnetism of this compound, will be helpful for the advancement of the knowledge of field-induced first-order magnetic transitions.

We would like to thank Kartik K Iyer for his help during this investigation.

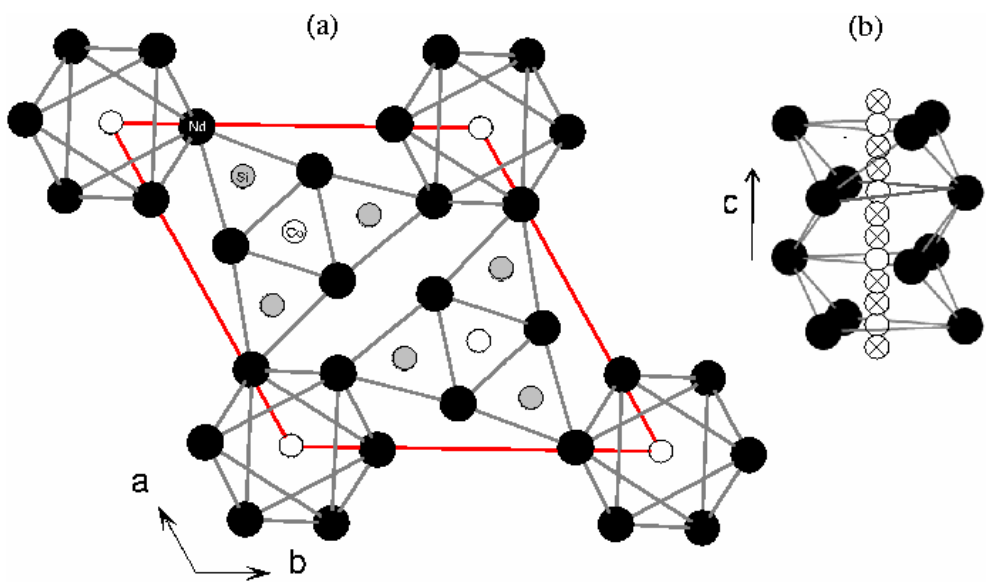

Figure 1:

The crystal structure of $Nd_6Co_{1.67}Si_3$ projected along the c-axis is shown in **(a)**. The arrangement of face-shared Nd octahedra running along **c**-axis connecting the Co and Si centered chains (see **(a)**) is shown in **(b);** the open circle corresponds to Co atoms and the Co positions (with cross marks inside the circles) within the Nd octahedra are randomly filled. Nd atoms are represented by filled black (big) circles, whereas Si atoms are marked by grey circles.

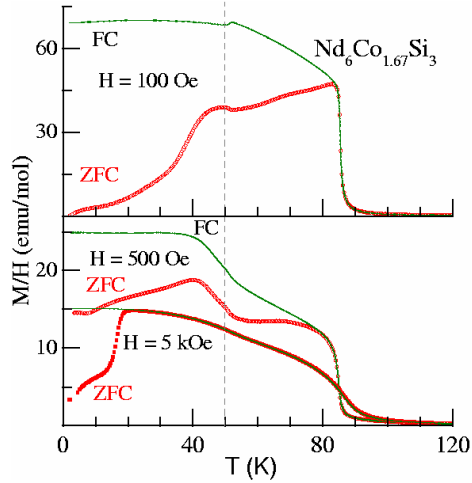

Figure 2:
(color online) Dc magnetization divided by magnetic field as a function of temperature, measured in the presence of 100 Oe, 500 Oe, and 5 kOe, for $Nd_6Co_{1.67}Si_3$, for the zero-field-cooled and field-cooled conditions of the specimen. A vertical dashed line is drawn to highlight the existence of an anomaly around 50K.

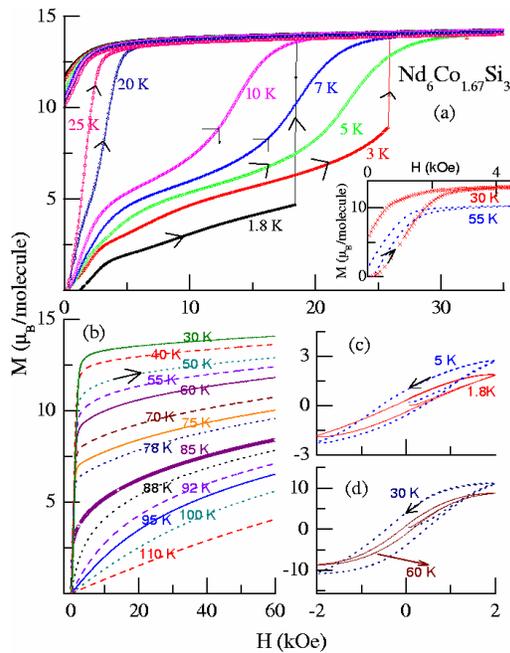

Figure 3:
(color online) **(a)** and **(b)** Isothermal dc magnetization behavior at several temperatures for the zero-field-cooled condition of the specimen, $Nd_6Co_{1.67}Si_3$, with the measurements extended to high fields as described in the text. In **(a),** the data for both 'up' and 'down' field directions are shown and, in the inset, the curves for 30 and 55 K are shown to highlight how these evolve at low fields with decreasing temperature. In **(c)** and **(d),** the low-field hysteresis loops (-2 to +2 kOe) at selected temperatures are shown.

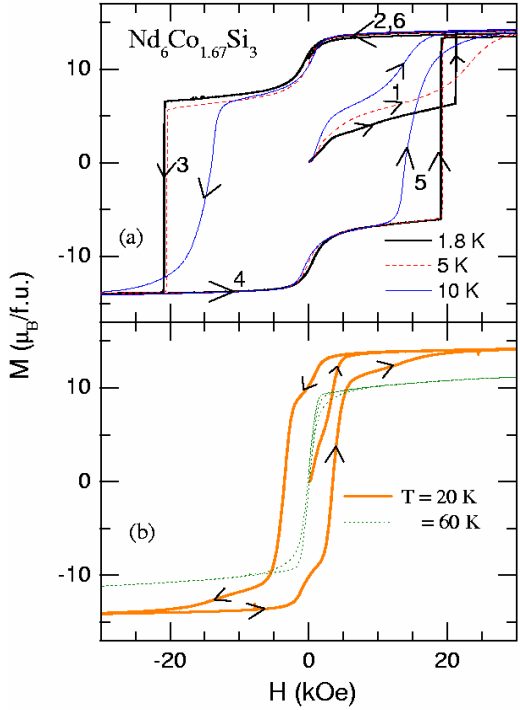

Figure 4:
(color online) Hysteresis loop behavior of isothermal magnetization **(a)** at 1.8, 5 and 10 K and **(b)** 20 and 60 K for the zero-field-cooled condition of the specimen, $Nd_6Co_{1.67}Si_3$, for a variation of the field from 0 to 30 to -30 to 0 kOe. The numericals 1 to 6 represent paths for the variation of the field. Note that, in **(a)**, except virgin part (along path 1), the envelope curves overlap for 1.8 and 5 K.

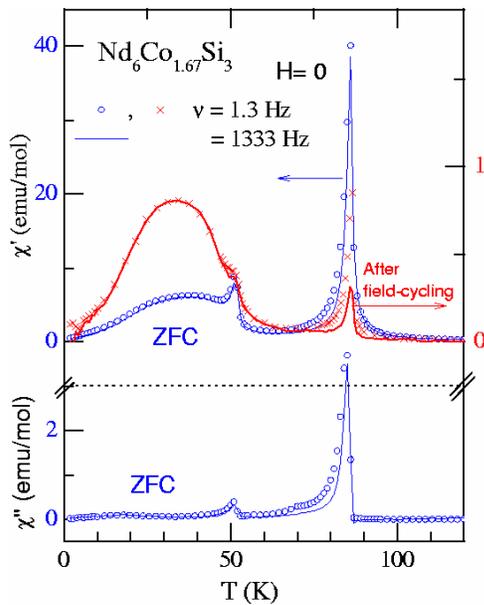

Figure 5:
(color online) Real and imaginary parts of ac magnetic susceptibility as a function of temperatures for the zero-field-cooled condition of the specimen, $Nd_6Co_{1.67}Si_3$, for two frequencies. The real part of the curve obtained in zero-field after a high-field cycling is also shown and the intensity is considerably reduced. The data points are shown for 1.3 Hz only, whereas for 1333 Hz, the lines obtained by joining the data points are shown for the sake of clarity.

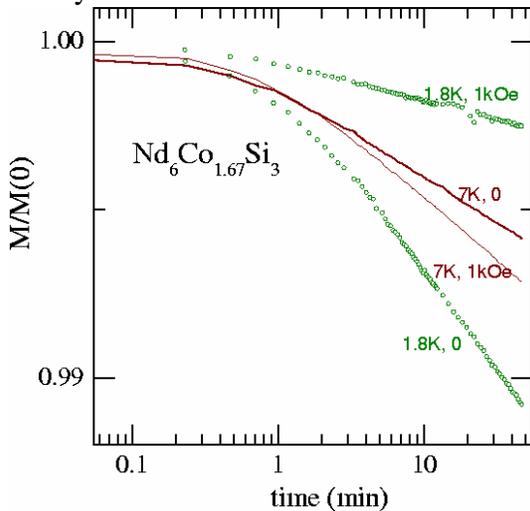

Figure 6:
(color online) The magnetic relaxation behavior obtained in 0 and 1 kOe after a field cycling (see text) at 1.8 and 7 K for $Nd_6Co_{1.67}Si_3$. M(0) refers to the initial value after attaining the field specified. For the sake of clarity, for two curves, the lines obtained by joining data points are only shown.

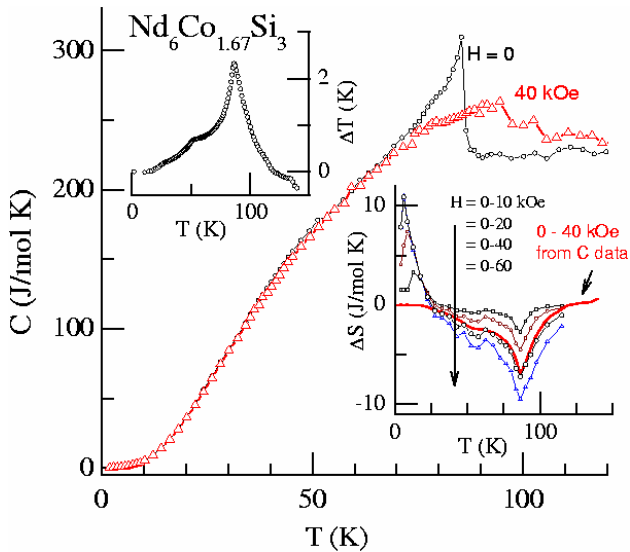

Figure 7:
(color online) Heat capacity as a function of temperature in zero field and in the presence of 40 kOe for $Nd_6Co_{1.67}Si_3$. A line is drawn through the data points in each curve. In the left inset, the adiabatic temperature change derived from this data is plotted. In the right inset, the isothermal entropy change derived from magnetization are plotted for four final fields, along with the corresponding data for one field from C(T). The lines through the data points serve as guides to the eyes.